\documentclass[twocolumn,showpacs,preprintnumbers,amsmath,amssymb,epsfig,floatfix,prb]{revtex4-2}
\usepackage[utf8]{inputenc}
\usepackage{siunitx}
\usepackage{tikz}
\usetikzlibrary{positioning}
\usepackage{amsmath}
\usepackage{pifont}
\usepackage{mathtools}
\usepackage{amsfonts}
\usepackage{amssymb}
\usepackage{graphicx,stackengine}
\usepackage{physics}
\usepackage[version=3]{mhchem}
\usepackage{xcolor}
\usepackage[english]{babel}
\usepackage{subfigure}
\usepackage[hidelinks,
colorlinks=true,
linkcolor=blue,
urlcolor=blue,
citecolor=blue]{hyperref}
\usepackage{calligra}
\usepackage{calrsfs}
\DeclareMathAlphabet{\mathcalligra}{T1}{calligra}{l}{m}
\DeclareMathAlphabet{\altmathcal}{OMS}{cmsy}{m}{n}
\usepackage{epsfig}
\usepackage{color}
\usepackage{natbib}
\usepackage[english]{babel}
\usepackage{breakurl}
\hypersetup{colorlinks=true, citecolor=blue, urlcolor=blue, linkcolor=blue}
\usepackage{bm}
\usepackage{gensymb}
\usepackage{color, colortbl}
\usepackage{overpic,color}
\usepackage{booktabs}
\definecolor{indigo}{rgb}{0.30,0.00,0.50}

\newcolumntype{C}{>{$}c<{$}}
\newcolumntype{L}[1]{>{\raggedright\arraybackslash}p{#1}}
\newcolumntype{C}[1]{>{\centering\arraybackslash}p{#1}}
\newcolumntype{R}[1]{>{\raggedleft\arraybackslash}p{#1}}

\begin{document}
\title{High temperature antiferromagnetism in ultrathin \ce{SrRu2O6} nanosheets}
\author{Deepak K. Roy}
\affiliation{Department of Physics, Indian Institute of Science Education and Research, Pune 411008, India}
\author{Mukul Kabir}
\email{mukul.kabir@iiserpune.ac.in} 
\affiliation{Department of Physics, Indian Institute of Science Education and Research, Pune 411008, India}

\date{\today}

\begin{abstract}
The quest for room-temperature nanoscale magnets remains a central challenge, driven by their promising applications in quantum technologies. Layered $4d$ and $5d$ transition metal oxides with high magnetic ordering temperatures offer significant potential in this context. We explore ultrathin \ce{SrRu2O6} nanosheets using first-principles calculations, complemented by the classical Heisenberg Monte Carlo simulations. Remarkably, these nanosheets exhibit robust antiferromagnetic ordering with N\'eel temperatures exceeding 430 K, despite the enhanced spin fluctuations characteristic of two-dimensional systems. Surface-termination-induced intrinsic charge doping introduces complexity to the magnetism, resulting in an insulator-to-metal transition and renormalized N\'eel temperatures in doped systems. A detailed microscopic analysis reveals distinct mechanisms underlying the magnetic behavior in electron- and hole-doped nanosheets. These findings provide a foundation for advancing theoretical and experimental studies in the largely unexplored realm of correlated oxides at the two-dimensional limit.
\end{abstract}
\maketitle

%

\section{Introduction}
Ruthenates present a rich platform for exploring exotic physics owing to the intricate interplay between competing electronic correlations, spin-orbit (SO) coupling and the dispersion of Ru-$4d$ orbitals, which governs the electron kinetic energy. The Ruddlesden-Popper series of ruthenate perovskites \ce{(Ca/Sr)_{$n$+1}Ru_nO_{3$n$+1}}, display a diverse range of intriguing phenomena.~\citep{Physics_Uspekhi46_1_2003_Ovchinnikov,MSEB63_76_82_1999_Cao}  For instance, \ce{Sr2RuO4} ($n=1$) is among the first non-copper-based materials to exhibit triplet superconductivity.~\citep{Nature372_532_1994_Maeno,APL60_1138_1992_Lichtenberg,MRB35_11_2000_Mao} 
In the $n = 2$ phase, \ce{Sr3Ru2O7} exhibits borderline ferromagnetism (FM) with non-Fermi liquid characteristics, while \ce{Ca3Ru2O7} is a metallic A-type antiferromagnet (AFM) over a narrow temperature range.~\citep{PhysRevLett.86.2661_Perry,PNAS117_6_2852_Mousatov,PhysRevB.63.165101_DJ_Singh,PhysRevLett.78.1751_Cao,PhysRevB.84.014422_XKe,NatPhy15_671_2019_Sokolov} At low temperatures,  \ce{Sr4Ru3O10} ($n=3$) features a unique FM configuration with alternating in-plane and out-of-plane \ce{Ru}-spin orientations.~\citep{JMMM493_165698_2020_Capogna,PhysRevB.98.064418_Zheng} The infinite-layer \ce{SrRuO3} ($n=\infty$) is an itinerant ferromagnet with intriguing transport properties, transitioning from a Fermi liquid state at low temperatures to a bad metal state at higher temperatures.~\citep{1.1656282,RevModPhys.84.253_Koster,PhysRevX.9.011027} In the realm of exotic quantum phases, $\alpha$-\ce{RuCl3} exhibits behaviour characteristic of a proximate Kitaev spin liquid, featuring fractional excitations.~\citep{PhysRevB.91.144420_Sears,NatPhy13_1079_2017_SHDo,s41567-018-0129-5,PhysRevB.99.094415}

While most ruthenates exhibit the \ce{Ru^{4+}} oxidation state, the \ce{Ru^{5+}} oxides reveal intriguing properties.  For example, \ce{SrRu2O6} is an antiferromagnet with an exceptionally high N\'eel temperature, $T_{\rm N} \sim$ 560 K.~\citep{AngewChem53_4423_2014_Hiley,PhysRevB.92.104413_Hiley,PhysRevB.92.100404_Tian,acs.chemmater.0c02469} Unlike the cubic lattice and corner-sharing \ce{RuO6} octahedra in perovskites, \ce{SrRu2O6} adopts 
a structure with edge-shared octahedra arranged in a honeycomb \ce{Ru} lattice.  This configuration features dual \ce{Ru-O-Ru} superexchange pathways and a \ce{Ru^5+}-$4d^3$ electronic configuration. Such high-temperature antiferromagnets are rare in compounds lacking 3$d$ transition elements, with notable exceptions such as \ce{SrTcO3}, \ce{CaTcO3}, and \ce{NaOsO3}.~\citep{PhysRevLett.106.067201_Rodriguez,JACS133_6_1654_2011_Avdeev,PhysRevB.80.161104}. Compared to these perovskites, \ce{SrRu2O6} is distinctive due to its quasi-two-dimensional magnetic structure, where magnetic \ce{Ru2O6} sheets are separated by \ce{Sr} layers along the perpendicular direction. Despite weak interlayer coupling, neglected in some theotrtical studies,~\citep{PhysRevB.92.104413_Hiley,NatMat18_563_2019_Suzuki} the material exhibits an unexpectedly high ordering temperature. Another remarkable feature is its reduced magnetic moment of about 1.4$\mu_{\rm B}$,~\citep{PhysRevB.92.104413_Hiley,PhysRevB.92.100404_Tian} significantly smaller than the expected 3$\mu_{\rm B}$ for a half-filled $t_{2g}$ configuration. This reduction is attributed to strong covalency between Ru-$d$ and O-$p$ orbitals. Furthermore, while \ce{SrRu2O6} is classified as a band insulator, the underlying transport mechanism and the role of defects remain subjects of ongoing debate.~\citep{PhysRevB.92.134408_Streltsov,PhysRevB.96.155135_Atsushi_Hariki,NatMat18_563_2019_Suzuki,SciRep7_11742_2017_Okamoto,PhysRevB.108.195137_Csontosova}

Contrary to the electronic structure, the magnetic properties of bulk \ce{SrRu2O6} are well characterized.~\citep{AngewChem53_4423_2014_Hiley,PhysRevB.92.104413_Hiley,PhysRevB.92.100404_Tian,acs.chemmater.0c02469} However, an intriguing question arises regarding the evolution of its high-temperature magnetism as the material is thinned to the monolayer limit, particularly in light of the growing interest in two-dimensional (2D) magnetism. It would be fascinating to determine whether high-temperature antiferromagnetism persists in the 2D regime. The recent hydrothermal synthesis of ultrathin \ce{SrRu2O6} flakes has sparked further interest in exploring the magnetic properties of 4$d$ oxides in their 2D form. The exfoliation of chemically bonded compounds is challenging and rarely achieved,~\citep{AdvFunctMat_27_1603254_2017_FWang,s41565-018-0134-y,AdvMatInterface5_1800549_2018_TPYadav,ChemMat30_17_5923_2018_APBalan}  unlike the van der Waals (vdW) materials. While most non-vdW metal-oxide magnets consist of 3$d$ transition metals,~\citep{s41565-018-0134-y,AdvMatInterface5_1800549_2018_TPYadav,ChemMat30_17_5923_2018_APBalan}  2D ruthenates are particularly intriguing due to the complex interplay between electron correlations, SO coupling, and electronic bandwidth. Moreover, the surfaces of non-vdW oxides often undergo reconstruction due to the presence of dangling bonds, which can significantly impact their properties. The nature of the surface termination can disrupt charge neutrality, leading to inherent carrier doping. This combination of surface reconstruction and natural doping introduces additional layers of complexity that can be further exploited to realize novel properties.~\citep{Roy_PhysRevB.110.L020403} In this study, we investigate the evolution of magnetism in \ce{SrRu2O6} as the layer thickness is reduced to the 2D limit and examine the impact of carrier doping on its magnetic properties.  

From a technological perspective, antiferromagnetic spintronics offers significant advantages over traditional ferromagnetic systems.~\citep{NatNanotech11_231_2016_Jungwirth,RevModPhys.90.015005_VBaltz,s41563-023-01492-6} The absence of stray fields, ultrafast spin dynamics, and the potential for electrical manipulation render AFM materials highly promising for next-generation technologies. In this context, 2D materials and their heterostructures provide an ideal platform for exploring novel magnetic properties. While ferromagnetic 2D materials have garnered significant attention, antiferromagnetic 2D materials have not elicited the same level of excitement.~\citep{acs.nanolett.6b03052,acsnano.1c06864,acsnano.1c09150,Olsen_2024} Nonetheless, the technological relevance of 2D magnetism hinges on achieving magnetic order above room temperature, a challenge that has proven difficult to overcome.~\citep{s41586-018-0626-9,PhysRevB.103.214411,PhysRevMaterials.6.084407} In this context, ultrathin oxide flakes that exhibit high-temperature antiferromagnetism in their bulk form present significant potential. This study explores the microscopic magnetic exchange interactions and demonstrates that ultrathin \ce{SrRu2O6} flakes retain room-temperature antiferromagnetism. This finding paves the way for further experimental investigations and potential applications.

\section{Methodology}
First-principles calculations are conducted using density functional theory,~\citep{PhysRev.136.B864,PhysRev.140.A1133} as implemented in the Vienna {\em ab initio} simulation package.~\citep{PhysRevB.48.13115,PhysRevB.54.11169,doi:10.1063/1.2187006} The wave functions are represented using the projector-augmented wave formalism,~\citep{PhysRevB.50.17953} with a plane-wave basis set energy cut-off of 600 eV. The exchange-correlation energy is treated using the Perdew-Burke-Ernzerhof functional.~\citep{PhysRevLett.77.3865}  To make a comparison, select calculations are also performed with the range-separated Heyd–Scuseria–Ernzerhof (HSE06) hybrid functional~\citep{HSE_functional} and the recently developed $r^2$SCAN meta-GGA functional.~\citep{R2SCAN_functional} The bulk \ce{SrRu2O6} is modelled using a $1 \times 2 \times 2$ supercell to capture different magnetic configurations, including antiferromagnetic interlayer coupling. This approach allows for a comprehensive analysis of the magnetic interactions. For ultrathin nanosheets, a vacuum space of 20 \si{\angstrom} is incorporated to minimize interactions between periodic images. A $\Gamma$-centered $8 \times 4 \times 4$ Monkhorst-Pack $k$-mesh is employed for the bulk calculations,~\citep{PhysRevB.13.5188} while a finer $10 \times 5 \times 1$ $k$-mesh is utilized for the ultrathin nanosheets to ensure accurate sampling of the Brillouin zone in both cases. Although the Hubbard-type on-site Coulomb interaction  $U$  is not explicitly included in our primary calculations to describe the \ce{Ru}-$4d$ electrons, selected computations are repeated with \( U_\text{Ru} \leqslant 1.8 \, \si{\electronvolt} \) to assess its potential effects.~\citep{PhysRevB.57.1505} The lattice parameters and ionic positions for both the bulk and nanosheets, across various magnetic configurations, are optimized until all force components are reduced below a threshold of 0.01 \si{\electronvolt/\angstrom}.  Although ultrathin flakes have been experimentally synthesized, we assess their dynamic stability by calculating the phonon band structure using the finite difference method. Larger $2\cross4\cross1$ supercells are employed, and the force constants are derived using the Phonopy code.~\citep{Phonopy_scripta_materialia} A stricter convergence criterion with force components reduced below 0.001 \si{\electronvolt/\angstrom}, along with an increased kinetic energy cut-off of 700 \si{\electronvolt}, is employed for phonon calculations. The magnetism is modelled using the classical Heisenberg Hamiltonian, with exchange parameters derived from first-principles calculations. Magnetic ordering temperatures are determined via Monte Carlo simulations using the Vampire Code,~\citep{Vampire_JPCM_Evans} where over $10^4$ spins are included to minimize finite-size effects. At each temperature, 1.5$\times$10$^6$ Monte Carlo steps are employed to ensure thermal equilibrium.

\section{Results and Discussion}
We start by analyzing the bulk properties of \ce{SrRu2O6}, comparing the results with both experimental data and previous theoretical results. Subsequently, we employ the same theoretical framework to explore ultrathin flakes and investigate the effect of charge carrier doping. 

\subsection{Insulating AFM bulk}
Bulk \ce{SrRu2O6} crystallizes in the trigonal $P\bar{3}1m$ phase, with calculated parameters of $a$ = 5.31 \AA, and  $c$ = 5.20 \AA, which are in good agreement with experimental data.~\cite{PhysRevB.92.100404_Tian,PhysRevB.92.104413_Hiley} The crystal structure consists of quasi-2D layers of edge-sharing \ce{RuO6} octahedra, forming a honeycomb \ce{Ru}-lattice in the $ab$-plane. These magnetic layers are stacked along the $c$-axis, separated by alternating \ce{Sr} layers.

\begin{figure}[!t]
\includegraphics[width=0.95\linewidth]{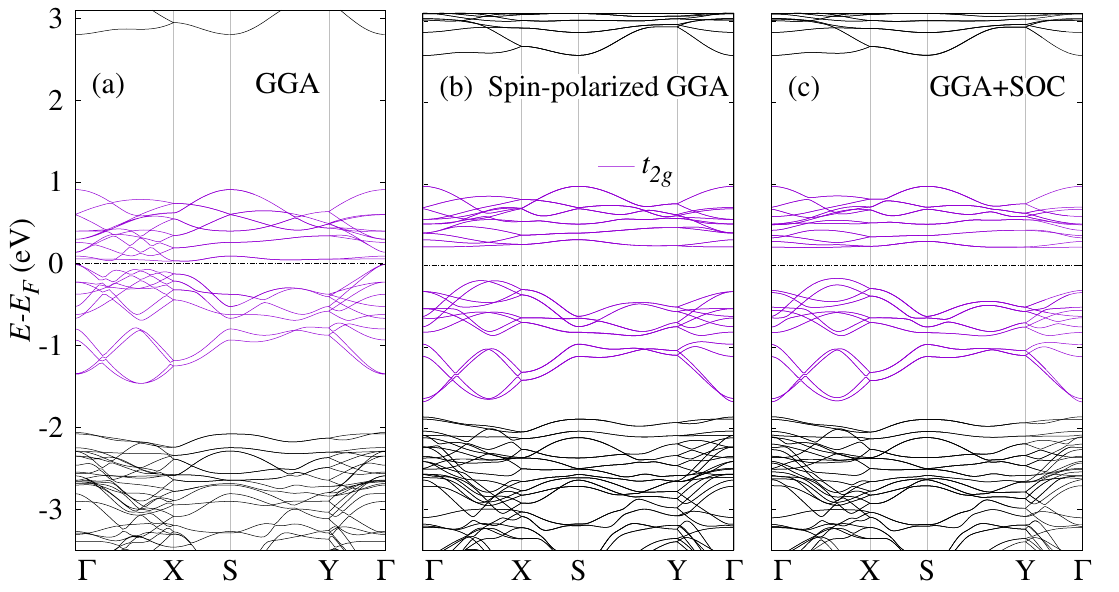}
\caption{The electronic band structure of bulk \ce{SrRu2O6} at various theoretical hierarchy reveals key features. The trigonal structural distortion splits the $t_{2g}$ manifold into an $a_{1g}$ singlet and  $e'_g$ doublets. A non-magnetic gap of 40 meV, as shown in (a), indicates the band-insulating nature. (b) When magnetism is incorporated, the gap increases, but SO coupling has a negligible effect, as demonstrated in (c). The unoccupied $e_g$ bands appear approximately 2.5 \si{\electronvolt} above the Fermi level.
}
\label{fig:Fig.2}
\end{figure}

The electronic structure of bulk \ce{SrRu2O6} remains widely debated and is not the primary focus of this study. Using the PBE-GGA exchange-correlation functional,~\citep{PhysRevLett.77.3865} the non-magnetic bulk \ce{SrRu2O6} reveals a small band gap of 40 meV at the Fermi level, indicating a normal insulating state [Figure~\ref{fig:Fig.2}(a)]. Upon introducing magnetism, the band gap increases with the \ce{Ru} magnetic moment,~\citep{PhysRevB.92.134408_Streltsov} reaching 0.41 eV [Figure~\ref{fig:Fig.2}(b)]. The overall electronic structure remains largely unchanged as SO coupling is introduced, though the energy gap is slightly reduced by 30 meV [Figure~\ref{fig:Fig.2}(c)], corroborating earlier calculations.~\citep{PhysRevB.91.214420_DJ_Singh_SrRu2O6,PhysRevB.92.134408_Streltsov,PhysRevB.96.155135_Atsushi_Hariki} Calculations were repeated with an added Hubbard correlation $U_{\rm \ce{Ru}}$ between 1.3 and 1.8 eV, resulting in a significantly larger gap, ranging from 0.86 to 1.03 eV. Additionally, advanced exchange-correlation functionals like $r^2$SCAN meta-GGA~\citep{R2SCAN_functional} and hybrid HSE functionals~\citep{HSE_functional} produced higher gaps of 0.87 eV and 2.5 eV, respectively. This behaviour aligns with that of a band insulator, where an existing non-magnetic gap is enhanced by magnetism, electron correlation, and relativistic effects. However, resistivity measurements in the high-temperature range indicate an activated gap of 36 meV, significantly smaller than the theoretical gap predicted in this study and in previous reports.~\citep{PhysRevB.92.104413_Hiley,PhysRevB.92.100404_Tian,PhysRevB.91.214420_DJ_Singh_SrRu2O6,PhysRevB.92.134408_Streltsov,PhysRevB.96.155135_Atsushi_Hariki,SciRep7_11742_2017_Okamoto} The electron transport mechanism over a wide temperature range remains unclear, but it is suggested that the unusually small energy gap may arise from disorder-induced localized states in experimental samples. Further theoretical and experimental investigations are necessary to resolve this discrepancy.

The nominal valence of \ce{Ru^{5+}} in a large octahedral crystal field,  $\Delta_{\rm OC} \sim 3$ eV, leads to half-filled 4$d$-$t_{2g}^3$ orbitals, which exhibit a broad dispersion exceeding 2.5 eV. Trigonal compression of the \ce{RuO6} octahedra along the $c$-axis, manifested by the \ce{O-Ru-O} bond angle of 94.4$\degree$, further splits the $t_{2g}$ manifold into a singlet $a_{1g}$ and a doubly degenerate $e’_{g}$ by $\Delta_t \sim 0.6$ eV [Figure~\ref{fig:Fig.2}(b)]. These findings are consistent with resonant inelastic X-ray scattering (RIXS) data.~\citep{NatMat18_563_2019_Suzuki} The subbands are narrower, with widths of $W_{a_{1g}} \sim$ 0.74 eV and $W_{e'_{g}} \sim$ 0.68 eV [Figure~\ref{fig:Fig.2}(b)], and remain unaffected by the introduction of SO coupling [Figure~\ref{fig:Fig.2}(c)]. In contrast, the empty $t_{2g}$ states above the Fermi level are flatter with reduced widths of $W_{a_{1g}} \sim$ 0.36 eV and $W_{e'_{g}} \sim$ 0.46 eV,  suggesting lower kinetic energy, which supports the localized-itinerant picture discussed in the literature.~\citep{PhysRevB.92.134408_Streltsov,SciRep7_11742_2017_Okamoto,PhysRevB.96.155135_Atsushi_Hariki,PhysRevB.108.195137_Csontosova}. The effective correlation calculated from the occupied and unoccupied $t_{2g}$ states near the Fermi level is about 0.93 eV. This places \ce{SrRu2O6} near the fluctuating limit of $U \sim W$, similar to what has been argued for antiferromagnetic \ce{SrTcO3}, which exhibits high $T_{\rm N} \sim$ 1023 K.~\citep{PhysRevLett.106.067201_Rodriguez,PhysRevLett.108.197202_Mravlje}  Consequently, bulk \ce{SrRu2O6} is also expected to have a high ordering temperature. 

Consistent with experimental predictions,~\citep{PhysRevB.92.100404_Tian,PhysRevB.92.104413_Hiley} present calculations indicate a G-type AFM ordering in bulk \ce{SrRu2O6}, having both in-plane and out-of-plane AFM interactions. Within the edge-shared octahedral cage of \ce{O} atoms, with a \ce{Ru-O-Ru} bond angle of 103$\degree$, the \ce{Ru}-$t^3_{2g}$ electrons interact through competing processes that contribute to the exchange interactions.~\citep{PhysRevB.92.100404_Tian,PhysRevB.92.075112_Wang} The combined effects of direct exchange and superexchange mechanisms result in AFM interactions, in accordance with the Goodenough-Kanamori-Anderson (GKA) rules.~\citep{goodenough1963magnetism} Specifically, the nearest-neighbour superexchange is dominated by virtual $d_{xz}-d_{yz}$ and $d_{xy}-d_{xy}$ hoppings, while second-neighbour $d_{xz}-d_{yz}$ and $d_{xy}-d_{yz}$ hoppings, as well as third-neighbour $d_{xy}-d_{xy}$ hoppings, further contribute to the resultant AFM interaction. The other magnetic configurations, such as ferromagnetic (FM, 0.24 eV/\ce{Ru}), zigzag-AFM (0.15 eV/\ce{Ru}), and stripe-AFM (0.09 eV/\ce{Ru}), are energetically less favorable. This demonstrates relatively soft magnetism for a 4$d$ system, as indicated by the modest energy difference between the magnetic and non-magnetic states, which is 0.26 eV/\ce{Ru}. Furthermore, the \ce{Ru} spin moment, $m_{\rm s | Ru} = 1.36\mu_\text{B}$, is significantly lower than the expected value for a half-filled $t_{2g}$ ($S=3/2$) system. This reduction is attributed to the strong hybridization between \ce{Ru}-$d$ and \ce{O}-$p$ states, aligning with the experimentally observed ordered magnetic moment of 1.3--1.4$\mu_{\rm B}$.~\citep{PhysRevB.92.100404_Tian,PhysRevB.92.104413_Hiley} However, strong hybridization does not necessarily imply a weak coupling nature, as previously discussed. While the local moments in $t^3_{2g}$ perovskites are also reduced, the degree of reduction and the underlying physics differ across systems. For instance, \ce{SrMnO3} ($T_{\rm N} = 260$ K, $m_{\rm s | Mn} = 2.6 \mu_{\rm B}$) falls in the strong coupling regime with $U/W \gg 1$,~\citep{1974275} while \ce{NaOsO3} ($T_{\rm N} = 410$ K, $m_{\rm s | Os} = 1\mu_{\rm B}$) is in the weak coupling limit with $U/W \ll 1$.~\citep{PhysRevB.80.161104,PhysRevLett.108.257209} On the other hand, \ce{SrTcO3} ($T_{\rm N} = 1023$ K, $m_{\rm s | Tc} = 2.1\mu_{\rm B}$) resides at the boundary between these two extremes with $U/W \sim 1$.~\citep{PhysRevLett.106.067201_Rodriguez} Furthermore, the reduction in $m_{\rm s | Ru}$ is notably more severe compared to quasi-one-dimensional \ce{Li3RuO4}~\citep{PhysRevB.84.174430} and other mixed-metal oxides containing \ce{Ru^{5+}} cations.~\citep{BATTLE1984138,DARRIET1997139} The reduction in local moment observed in \ce{SrRu2O6} aligns with the interplay between electron correlation and bandwidth, characterized by $U \sim W$, resulting in significant hybridization between \ce{Ru}-$4d$ and \ce{O}-$2p$ orbitals. As expected in a half-filled $t^3_{2g}$ electronic configuration, the SO coupling in \ce{SrRu2O6}, with $\lambda < 200$ meV, does not play a dominant role in describing the electronic structure, resulting in vanishing orbital moment, $m_{\rm s | Ru} < 0.02 \mu_{\rm B}$. However, SO coupling plays a prime role in magnetism by inducing uniaxial spin anisotropy. 

\begin{figure*}[!t]
\includegraphics[width=0.9\linewidth]{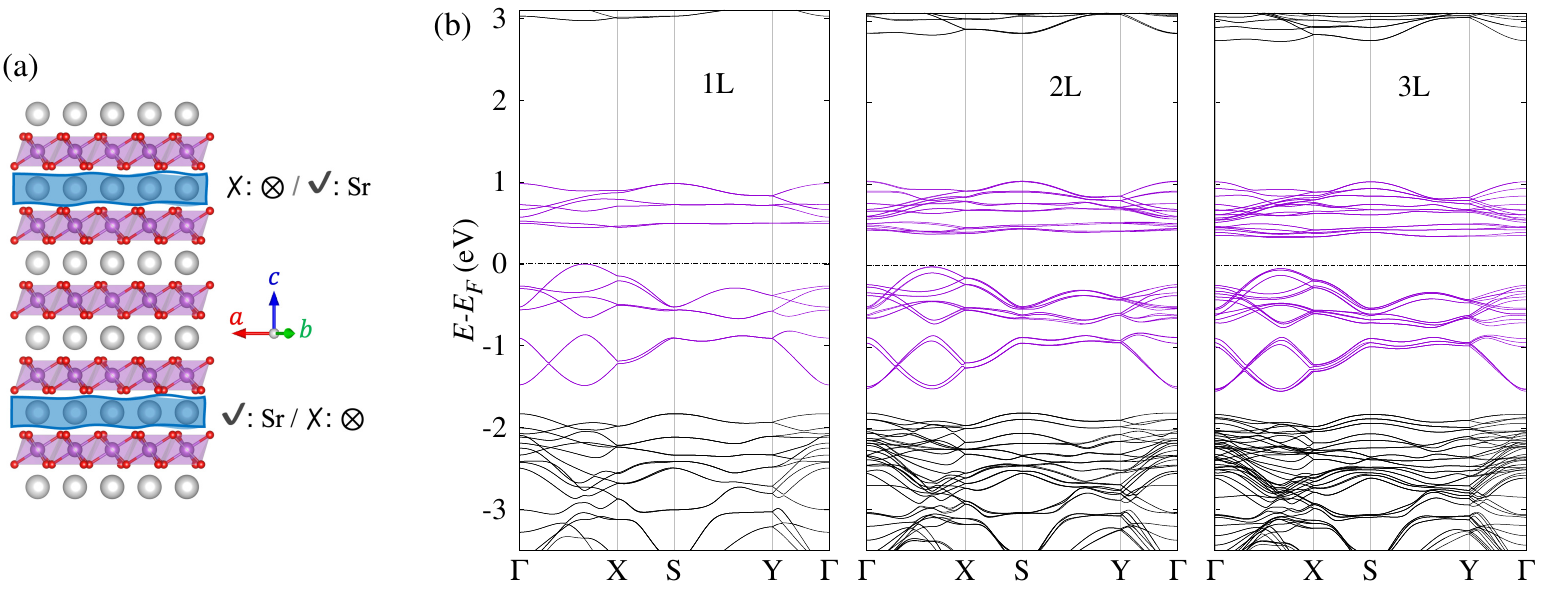}
\caption{(a) Bulk \ce{SrRu2O6} consists of two-dimensional \ce{Ru2O6} sublayers formed by edge-sharing octahedra within the $ab$-plane, stacked along the $c$-axis and separated by alternating \ce{Sr} layers. The \ce{Sr-O} bonds, not shown for clarity, are weaker and can be chemically cleaved, resulting in ultrathin flakes. Cleaving may occur at any \ce{Sr}-layer, highlighted in blue in the figure, leading to nanosheets with varying thicknesses. Flakes with (\ding{51}: \ce{Sr}) and without (\ding{55}: $\otimes$) \ce{Sr} termination have equal likelihood, giving rise to different types of flakes classified by their terminations. (b) Charge-neutral nanosheets are modelled by hydrogen passivating the surfaces of flakes terminated by \ce{Sr} layers on both sides, \ce{[H-Sr $|$$ \cdots$ $|$ Sr-H]}. The electronic structure of the one- to three-layer nanosheets generally mirrors that of the bulk, retaining their insulating nature. However, the reduced dimensionality introduces quantum confinement effects, which results in an increased band gap of 450 meV. 
}
\label{fig:BandNL}
\end{figure*} 
    
The electronic structure of the one- to three-layer \ce{SrRu2O6} nanosheets generally mirrors that of the bulk, retaining their insulating nature. However, the reduced dimensionality introduces quantum confinement effects, which result in a substantial increase in the band gap, reaching approximately 450 meV. This enhancement underscores the influence of confinement on the electronic properties of ultrathin layers, making them distinct from their bulk counterpart.  
    
To gain deeper insights into the microscopic nature of magnetism in bulk \ce{SrRu2O6} and ultrathin nanosheets, we employ the anisotropic Heisenberg model on the honeycomb \ce{Ru} lattice,~\citep{PhysRevB.103.214411,PhysRevMaterials.6.084407}
\begin{eqnarray}
\altmathcal{H} = &-& \frac{1}{2}\sum_{\mathclap{{\langle i, j\rangle}}} J_{1}\bm{S}_i \cdot \bm{S}_j  -   \frac{1}{2}\sum_{\mathclap{{\langle \langle i, j\rangle\rangle}}} J_{2}\bm{S}_i \cdot \bm{S}_j - \frac{1}{2}\sum_{\mathclap{{\langle \langle\langle i, j\rangle\rangle\rangle}}} J_{3}\bm{S}_i \cdot \bm{S}_j \nonumber \\ 
&-& \frac{1}{2}\sum_{\mathclap{{\langle i, j\rangle}}} J_\perp\bm{S}_i \cdot \bm{S}_j - \sum_i A_z S_i^z S_i^z. \nonumber
\end{eqnarray}
Here, $J_1$, $J_2$, and $J_3$ represent the first, second, and third nearest-neighbour intralayer isotropic exchange interactions, respectively, while $J_\perp$ denotes the interlayer interaction between \ce{Ru}-spins $\bm{S}_i$. The magnetic exchange coupling is FM for positive $J$ and negative for AFM interaction. Although SO coupling does not influence the electronic structure, it plays a crucial role in the magnetism of \ce{SrRu2O6} through on-site anisotropy, particularly in the ultrathin limit. The on-site anisotropy $A_z < 0$ characterizes in-plane and $A_z > 0$ represents out-of-plane magnetization. $A_z$ is calculated as the energy difference between the in-plane and out-of-plane magnetization, $A_z = (E_{[100]}-E_{[001]})/S^2$.  Such a magnetic model has been employed to describe 2D magnetic insulators, offering not only a quantitative framework but also the ability to predict the experimental evolution of magnetism under electrical gating.~\citep{PhysRevB.103.214411,PhysRevMaterials.6.084407} Isotropic interactions are calculated by energy mapping across various ordered phases, including FM, G-type AFM,  C-type AFM, in-plane zigzag-AFM, and in-plane stripe-AFM configurations. 
 
\begin{table*}[!t]
\caption{Heisenberg exchange interactions are derived from first-principles calculations. The magnetism in charge-neutral nanosheets is primarily governed by intralayer AFM $J_1$, AFM $J_3$, and out-of-plane magnetic anisotropy, resulting in high-temperature antiferromagnetic ordering in two dimensions, as calculated through Heisenberg Monte Carlo simulations. However, charge doping adversely affects the N\'eel temperature $T_\text{N}$. Electron doping significantly impacts the on-site anisotropy, flipping it to an in-plane orientation. In contrast, in hole-doped sublayers, both intralayer $J_2$ and $J_3$ transition to strong ferromagnetic interactions, accompanied by a substantial reduction in the \ce{Ru} spin moment, significantly influencing the magnetic ordering.
}
\begin{tabular}{L{4cm }R{1.3cm} R{1.3cm} R{1.3cm} R{1.3cm} R{1.1cm} C{1.1cm} C{1.1cm} R{1.1cm} C{1.2cm}}
\toprule
\toprule
\vspace{0.1cm}
                  & \multicolumn{4}{c} {Exchange interactions (meV)} & \multicolumn{1}{c}{$A_z$} & \multicolumn{1}{c}{$m_{\rm s | Ru}$} & \multicolumn{1}{c}{$m_{\rm o | Ru}$} & \multicolumn{1}{r}{$T_\text{N}$}\\
                  & \multicolumn{1}{c} {$J_1$}   &  \multicolumn{1}{c}{$J_2$}       &   \multicolumn{1}{c}{$J_3$}    &   \multicolumn{1}{c}{$J_\perp$} &  (meV) &  ($\mu_\text{B}$) &  ($\mu_\text{B}$) & \multicolumn{1}{r}{(K)} \\        
\midrule
Bulk  \ce{SrRu2O6}                               &  $-$163.2      &  $-$0.1  &   $-$12.2   & $-$0.85    & 3.5   &  1.36   & 0.02    & 501 \\

3L-[\ce{H-Sr}$|$$\cdots$$|$\ce{Sr-H}] & $-$163.1       &   $-$0.5  &  $-$10.0   &  $-$0.28   & 3.6   & 1.36    &  0.02    & 452 \\

2L-[\ce{H-Sr}$|$$\cdots$$|$\ce{Sr-H}] & $-$162.1       &   $-$0.8  &  $-$10.6   &  $-$0.14   & 3.7   & 1.36    &  0.02    & 445 \\

1L-[\ce{H-Sr}$|$\ce{Ru2O6}$|$\ce{Sr-H}] & $-$163.0    &   $-$1.8  &  $-$9.5     &  $-$    & 3.9   & 1.36    &  0.02    & 431 \\

1L-[\ce{H-Sr}$|$\ce{Ru2O6}$|$\ce{Sr}] & $-$136.0   &   $-$25.2  &  $-$9.7     &  $-$    & $-$3.8   & 1.03    &  0.04    & 73 \\
1L-[\ce{Sr}$|$\ce{Ru2O6}$|$\ce{Sr}]    & $-$90.0     &   4.6         &  $-$9.8     &  $-$    & $-$14.9   & 1.01    &  0.08    & 195 \\
1L-[\ce{H-Sr}$|$\ce{Ru2O6}$|$$\otimes$]    & $-$198.7     &   47.7         &  88.5     &  $-$    & 3.2   & 0.32    &  0.01    & 23 \\
2L-[$\otimes$$|$\ce{Ru2O6}$|$$\otimes$]    & $-$200.1     &   50.3         &  86.8     &   $-$10.2   & 3.7   & 0.32    &  0.01    & 29 \\
2L-[\ce{Sr}$|$\ce{Ru2O6}$|$\ce{Sr}]    & $-$131.6     &   $-$24.5         &  $-$13.9     &   0.22   & $-$3.8   & 1.02    &  0.04    & 91 \\
\bottomrule
\hline
\end{tabular}
\label{table1}
\end{table*}

Calculations reveal strong in-plane AFM correlations compared to the weaker out-of-plane interactions (Table~\ref{table1}). The robust in-plane magnetism is primarily governed by AFM superexchange coupling between \ce{Ru} spins, mediated by oxygen and further enhanced by the strong \ce{Ru-O} covalency. The honeycomb $J_1$-$J_2$-$J_3$ model provides an excellent fit, with the magnetism predominantly described by the first-neighbour AFM interaction, $J_1=163$ \si{\milli\electronvolt}, which corresponds to $J_1S^2 = 75$ \si{\milli\electronvolt}, and in good agreement with experimental estimates derived from magnon dispersion.~\citep{NatMat18_563_2019_Suzuki} While $J_3/J_1 = 0.07$ remains moderate, the second-neighbour interaction $J_2$ is negligible (Table~\ref{table1}). With a much weaker interplane exchange of $J_\perp = 0.85$  \si{\milli\electronvolt}, the magnetism in bulk \ce{SrRu2O6} is effectively 2D in nature. Calculated positive on-site anisotropy of $A_z=3.5$ \si{\milli\electronvolt} indicates an easy-axis magnetism, consistent with experimental predictions.~\citep{PhysRevB.92.100404_Tian,NatMat18_563_2019_Suzuki} Heisenberg Monte Carlo simulations, incorporating these exchange parameters, yield a N\'eel temperature of 501, which is in close agreement with the experimental value of 565 K.~\citep{AngewChem53_4423_2014_Hiley,PhysRevB.92.104413_Hiley,PhysRevB.92.100404_Tian,acs.chemmater.0c02469} 

Calculations for bulk \ce{SrRu2O6} are repeated using various descriptions of exchange-correlation functionals, but the results deviate significantly from experimental observations (Supplemental Material~\citep{supple}). For instance, the electronic band gap increases to 0.93 eV as electron correlation is supplemented with a Hubbard-like Coulomb interaction ($U_{\ce{Ru}}$=1.5 eV). A similar enhancement is observed with the regulated-restored $r^2$SCAN meta-GGA, and the hybrid HSE06 functional. While the G-type AFM ground state is well reproduced, these descriptions of exchange-correlation significantly overestimate the \ce{Ru} magnetic moment. Moreover, the calculated N\'eel temperatures are significantly lower than experimental values, mainly due to a reduction in the AFM $J_1$, resulting from a decreased energy difference between the G-AFM ground state and the stripy AFM excited state. 
  
\subsection{Two-dimensional \ce{SrRu2O6} sheets}
Having explored the bulk properties, it is intriguing to investigate \ce{SrRu2O6} sheets in the 2D limit [Figure~\ref{fig:BandNL}(a)]. Given the quasi-2D nature of magnetism in the bulk, the key question is whether these 2D sheets can sustain an above-room-temperature AFM state. The added complexities of quantum confinement, surface reconstruction, and self-doped charge carriers make this exploration more compelling. 

Bulk \ce{SrRu2O6} is a layered compound with alternating triangular \ce{Sr} layers and honeycomb \ce{Ru2O6} layers stacked along the $c$-axis [Figure~\ref{fig:BandNL}(a)]. We find that the \ce{Sr-O} bonds are about ten times weaker than the \ce{Ru-O} bonds. Consequently, the \ce{Sr-O} bonds are easily cleaved, leaving behind nanosheets of magnetic \ce{Ru2O6} layers in the $ab$-plane, separated by nonmagnetic \ce{Sr} layers. Since ultrathin sheets with or without \ce{Sr}-layer termination are equally likely, these 2D sheets become charge-doped at the surface. A deficiency or excess of \ce{Sr} from the stoichiometric composition results in hole or electron doping at the surface of the flakes, respectively. Driving the ultrathin sheets away from the charge neutral point could induce emergent physics, similar to observations in \ce{Na2IrO3} flakes.~\citep{Roy_PhysRevB.110.L020403} In the monolayer limit of \ce{Na2IrO3}, the Kitaev spin liquid state is disrupted, and carrier doping induces electronic and magnetic phase transitions, resulting in a spin-only ferromagnetic phase. 

The exfoliation energy for \ce{SrRu2O6} is calculated to be 76 meV/\AA$^2$, which is significantly lower than that of \ce{Na2IrO3}, where the \ce{Na-O} bonds are cleaved. Furthermore, this exfoliation energy is lower than that of metal carbides and nitrides, known as MXenes,~\citep{C7CP08645H,acsnano.9b06394,s44160-022-00104-6} which have already been experimentally cleaved, but is still higher than conventional vdW layered systems.~\citep{Exfoliation_vdw_PRB_2020} The calculated phonon dispersions show that both neutral and charge-doped ultrathin non-vdW flakes are dynamically stable (Supplemental Material~\citep{supple}). A flexural phonon mode with a small frequency is observed near the Brillouin zone centre, a feature commonly seen in vdW 2D materials and is attributed to numerical convergence.~\citep{Phonon_Zolyomi_PRB_2014,Phonon_Radescu_PRM_2019} These results are further corroborated by the recent successful synthesis of ultrathin \ce{SrRu2O6} nanosheets using a scalable liquid exfoliation technique.~\citep{ACS_ApplNanoMat4_9_9313_2021_Suvidya_SrRu2O6} Comprehensive structural characterizations were conducted using a variety of complementary experimental methods. The nanosheet thickness ranges from 1.3 to 2.2 \si{\nano\meter}, corresponding to three to five monolayers in the experiment. Nevertheless, the electronic and magnetic characterization of these nanosheets remains challenging, underscoring the critical importance of theoretical investigations. 

\subsection{Charge-neutral nanosheets}
\begin{figure}[!t]
\includegraphics[width=0.85\linewidth,trim={0 0 0 0},clip]{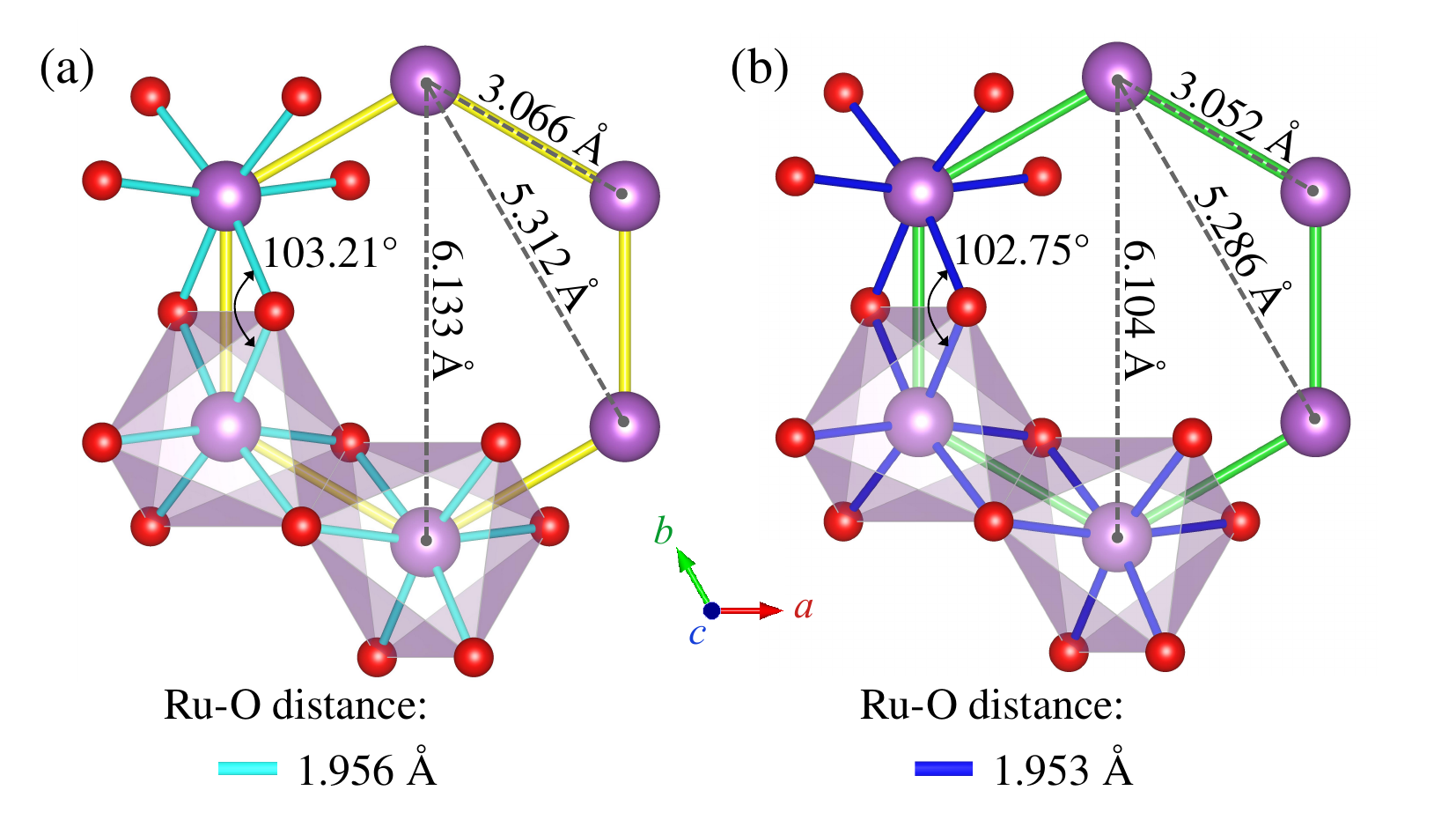}
\caption{
The structural comparison between (a) bulk \ce{SrRu2O6} and (b) the hydrogen-passivated charge-neutral monolayer reveals key differences and similarities. The honeycomb \ce{Ru} lattice and edge-sharing octahedral geometry remain evident in both cases. In the monolayer, a minor contraction of the in-plane lattice parameter is observed, impacting the \ce{Ru-Ru} distances. However, the \ce{Ru-O} bonds remain largely unaffected in the passivated nanosheets.
}
\label{fig:CompareStructure}
\end{figure}

We begin by investigating ultrathin layers comprising one to three monolayers, each terminated with \ce{Sr} layers on both sides and further hydrogen-passivated, forming \ce{[H-Sr $|$$ \cdots$ $|$ Sr-H]} structures. Bader charge analysis reveals that the magnetic honeycomb \ce{Ru2O6} sublayers in these nanosheets retain an identical charge distribution to that of bulk \ce{SrRu2O6}, confirming the charge neutrality of these flakes (Supplemental Material~\citep{supple}).  The in-plane lattice parameter shows minimal variation with thickness (Figure~\ref{fig:CompareStructure} and Supplemental Material~\citep{supple}), being only 0.5\% shorter in the monolayer compared to the bulk. As a result, the \ce{Ru-Ru} distances within the honeycomb lattice are also slightly reduced in the flakes. Other structural parameters remain unchanged, including \ce{Ru-O} bond distance and the degree of trigonal distortion.

The overall electronic structure of the nanosheets closely resembles that of the bulk, preserving the normal insulating state [Figure~\ref{fig:BandNL}(b)]. However, in the monolayer limit, the band gap increases to 450 \si{\milli\electronvolt}, a change attributed to quantum confinement effects rather than structural modifications, which are found to be negligible. Key electronic parameters, including $\Delta_\text{OC}$, $\Delta_t$, electron correlation, and bandwidths $W_{a_{1g}}$ and $W_{e'_{g}}$, remain unchanged in the nanosheets [Figure~\ref{fig:BandNL}(b)]. Consequently, both the spin and orbital moments remain unaltered in the ultrathin nanosheets.~\citep{supple} 

\begin{figure}[!t]
\includegraphics[width=0.85\linewidth,trim={0 0 0 0},clip]{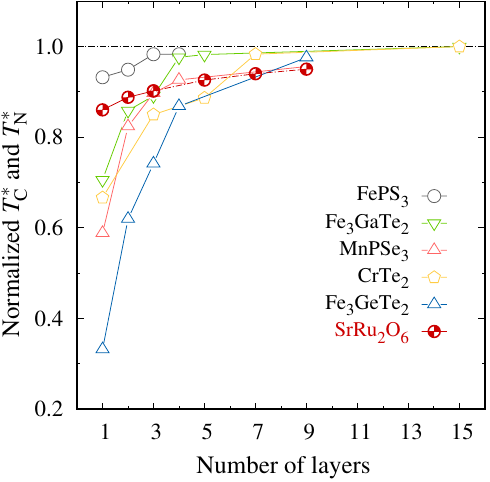}
\caption{The thickness-dependent evolution of magnetism in ultrathin \ce{SrRu2O6} nanosheets is compared to other 2D vdW magnets, with their ordering temperatures normalized to bulk values. The results for the passivated nanosheets calculated using the parameters in Table~\ref{table1} are compared with \ce{FePS3},~\citep{acs.nanolett.6b03052} \ce{Fe3GaTe2},~\citep{s41699-024-00460-1} \ce{MnPSe3},~\citep{s41565-021-00885-5} \ce{CrTe2},~\citep{s41467-021-22777-x} and \ce{Fe3GeTe2}.~\citep{s41586-018-0626-9} The $T_\text{N}$ for five- to nine-layer nanosheets (dotted line) are calculated using the interpolated exchange parameters.  Although magnetic ordering temperatures are renormalized in the 2D limit due to enhanced thermal spin fluctuations, the extent of this reduction varies significantly across materials, driven by differences in SO coupling and interlayer exchange interactions. 
}
\label{fig:Fig.Ordering}
\end{figure}

Remarkably, magnetism in \ce{SrRu2O6} 2D nanosheets remains robust, even at the monolayer limit, which consists of a single \ce{Ru2O6} sublayer. Similar to the bulk, the ultrathin films exhibit a G-type AFM structure, with the monolayer retaining the corresponding N\'eel AFM order. The retention of high-temperature magnetism in ultrathin flakes is surprising, considering the bulk exhibits quasi-2D magnetism with weak interlayer coupling $J_\perp$. Typically, the ordering temperaure undergoes significant renormalization in the 2D limit (Figure~\ref{fig:Fig.Ordering}), as observed in vdW magnets.~\citep{acs.nanolett.6b03052,s41699-024-00460-1,s41565-021-00885-5,s41467-021-22777-x,s41586-018-0626-9,acs.nanolett.9b05165} However, the renormalization of $T_\text{N}$ in \ce{SrRu2O6} nanosheets is significantly weaker (Figure~\ref{fig:Fig.Ordering}).

Similar to its bulk counterpart, magnetism in the nanosheets is primarily governed by AFM $J_1$ interactions and magnetic anisotropy oriented perpendicular to the honeycomb lattice (Table~\ref{table1}). Classical Heisenberg Monte Carlo simulations indicate a N\'eel temperature of 431 K for the passivated monolayer, making it the 2D antiferromagnet with the highest known ordering temperature. This surpasses not only experimentally synthesized materials~\citep{acs.nanolett.6b03052,acsnano.1c06864,acsnano.1c09150} but also those identified through computational screening.~\citep{Olsen_2024} In comparison, ternary \ce{MPS3} compounds, which share the $P\bar{3}1m$ space group, exhibit significantly lower ordering temperatures. For instance, \ce{FePS3} demonstrates Ising antiferromagnetism with an in-plane zigzag magnetic order below 118 K,\citep{acs.nanolett.6b03052} while \ce{MnPS3} is a Heisenberg antiferromagnet that exhibits out-of-plane N\'eel order below 78 K.\citep{acs.nanolett.9b05165} Compared to the bulk, the moderate reduction in the monolayer, with $T_\text{N}^\text{ML} = 0.86T_\text{N}^\text{bulk}$ is primarily attributed to the complete loss of magnetic interactions in the perpendicular direction, resulting in enhanced spin fluctuations. Additionally, a slight increase in the AFM $J_2$ further contributes to the decrease in $T_N$, as AFM $J_2$ hinders N\'eel ordering on a honeycomb lattice. The interlayer coupling $J_\perp$ remains AFM in ultrathin sheets, and it increases with layer thickness. This results in a moderate increase in the $T_\text{N}$ in the two- and three-layer nanosheets (Table~\ref{table1} and Figure~\ref{fig:Fig.Ordering})

\begin{figure*}[!t]
\includegraphics[width=0.90\linewidth]{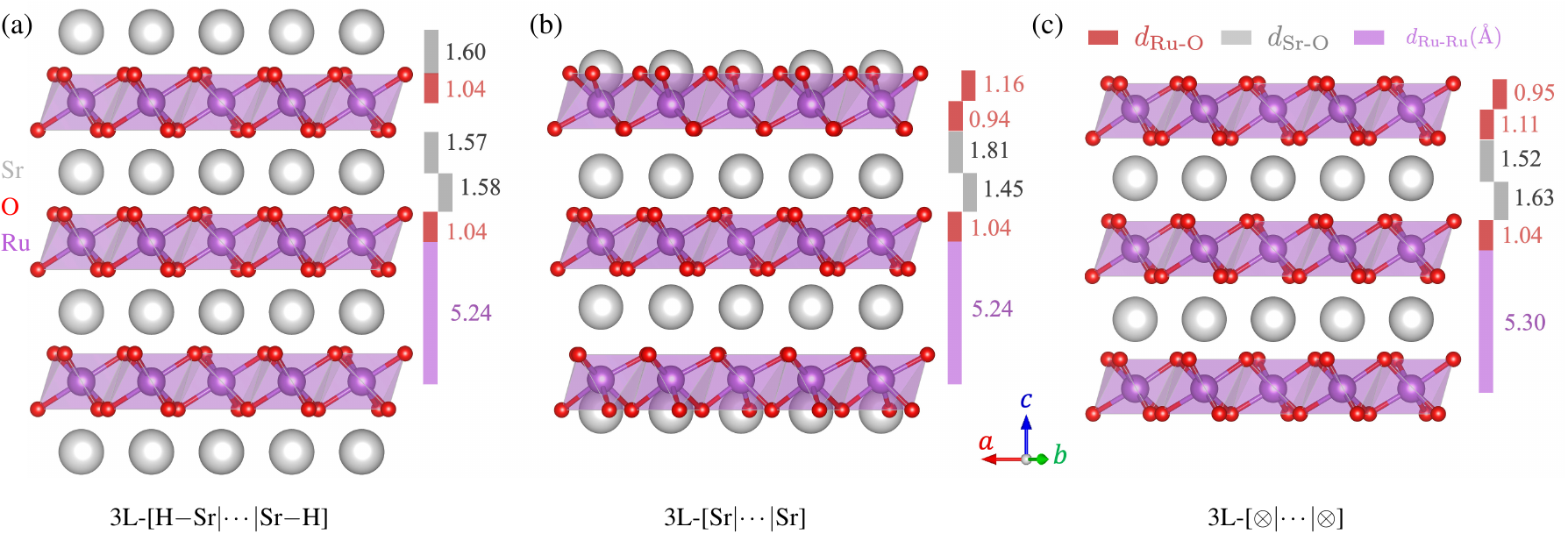}
\caption{The type of surface termination determines the charge carriers infused into the adjacent \ce{Ru2O6} sublayer, which in turn drives structural reconstruction and significantly influences the electronic and magnetic properties of the nanosheets. For instance, a comparison of structural reconstructions in three-layer nanosheets highlights these effects. (a) The passivated, charge-neutral nanosheet 3L-[\ce{H-Sr}$|$$\cdots$$|$\ce{Sr-H}] closely resembles the bulk structure without significantly altering the interplanar distances, as indicated by the colour bars. Passivating \ce{H}-atoms are not shown for clarity. (b) However, in the electron-doped 3L-[\ce{Sr}$|$$\cdots$$|$\ce{Sr}] flake, the exposed \ce{Sr} layer induces severe structural reconstruction, with the dangling \ce{Sr} layer plunging into the adjacent \ce{Ru2O6} sublayer. This results in significant changes to the interplanar distances, particularly $d_\text{\ce{Sr-O}}$ and $d_\text{\ce{Ru-O}}$. (c) In contrast, the structural reconstruction is relatively less pronounced in the hole-doped trilayer [$\otimes$$|$$\cdots$$|$$\otimes$] nanosheet. 
}
\label{fig:reconstruction}
\end{figure*}

\subsection{Charge-carrier doped nanosheets}
The non-vdW flakes become charge-doped depending on surface termination, which may severely impact their electronic and magnetic properties, as reported for \ce{Na2IrO3} nanosheets.~\citep{Roy_PhysRevB.110.L020403}  In the case of \ce{SrRu2O6}, the \ce{Sr}-terminated layer electron-dopes the adjacent \ce{Ru2O6} magnetic sublayer. In contrast, the oxygen-terminated layer hole-dopes it in the absence of a \ce{Sr} sublayer. Further, partial passivation allows for control over carrier density in the active magnetic layer.   

A clear trend in the in-plane lattice parameter $a$ is observed in the ultrathin flakes, independent of their thickness (Supplemental Material~\citep{supple}). The calculated $a$ increases in electron-doped flakes relative to their charge-neutral counterparts, whereas it decreases in hole-doped flakes. A concomitant trend is observed in structural parameters, such as the \ce{Ru-O} and \ce{Ru-Ru} bond distances. Structural reconstruction at the \ce{Sr}-terminated layer is pronounced and extends into the underlying \ce{Ru2O6} sublayer, as illustrated in Figure~\ref{fig:reconstruction} for nanosheets comprising three magnetic sublayers. As a result, nanosheets with and without the \ce{Sr} termination exhibit equivalent thickness. For instance, trilayer nanosheets have a thickness of approximately 1.3 \si{\nano\meter} regardless of termination(Figure~\ref{fig:reconstruction}), which is in excellent agreement with experimental findings.~\citep{ACS_ApplNanoMat4_9_9313_2021_Suvidya_SrRu2O6} Moreover, charge carriers influence other structural parameters, including \ce{Sr-O} distances, as well as \ce{Ru-O-Ru} and \ce{O-Ru-O} bond angles.~\citep{supple}  

Carrier doping significantly alters the electronic structure, leading to an insulator-to-metal transition in the adjacent \ce{Ru2O6} sublayer, irrespective of the termination. While the fully passivated monolayer [\ce{H-Sr}$|$\ce{Ru2O6}$|$\ce{Sr-H}] retains insulating behavior, hole-doped nanosheets [\ce{H-Sr}$|$\ce{Ru2O6}$|$$\otimes$], [$\otimes$$|$\ce{Ru2O6}$|$$\otimes$] and electron-doped [\ce{H-Sr}$|$\ce{Ru2O6}$|$\ce{Sr}] flakes exhibit metallic characteristics. The sole exception arises when the monolayer is terminated with \ce{Sr} layers on both sides,  [\ce{Sr}$|$\ce{Ru2O6}$|$\ce{Sr}], which remains insulating with a band gap of 188 \si{\milli\electronvolt} as the unoccupied $a_{1g}$ band becomes completely filled. The insulator-to-metal transition persists in thicker flakes, where metallic and insulating \ce{Ru2O6} sublayers coexist simultaneously. For instance, in the trilayer configuration [$\otimes$$|$\ce{Ru2O6}$|$$\cdots$$|$\ce{Ru2O6}$|$\ce{Sr}], both surface \ce{Ru2O6} sublayers, irrespective of \ce{Sr}-termination, exhibit metallic behavior, while the central \ce{Ru2O6} sublayer remains insulating. Furthermore, the density of states analysis indicates that irrespective of the surface termination, the \ce{Sr}-layers between the magnetic \ce{Ru2O6} sublayers remain insulating, indicating 2D surface metallicity in these nanosheets.

Magnetism is retained in the atomically thin sheets even under charge doping (Table~\ref{table1}). Despite significant changes in structural and electronic properties, the \ce{Ru2O6} sublayers maintain robust N\'eel AFM ordering. The interlayer exchange interaction remains antiferromagnetic in all cases, except in the bilayer with \ce{Sr} termination on both sides (Table~\ref{table1}), where the coupling becomes ferromagnetic between the two electron-doped metallic \ce{Ru2O6} sublayers (Table~\ref{table1}). Although magnetism persists, charge doping modifies the local \ce{Ru} spin and orbital moments, SO coupling, magnetic anisotropy, and exchange interactions. Several general trends are observed depending on the type and density of doping (Table~\ref{table1}).

First, charge carrier doping reduces the spin moment from 1.36$\mu_\text{B}$ in the charge-neural state, with the reduction being significantly more pronounced in hole-doped \ce{Ru2O6} sublayers ($m_{\rm s | Ru} \sim 0.3\mu_\text{B}$) compared to electron-doped ones ($m_{\rm s | Ru} \sim 1\mu_\text{B}$), despite identical carrier densities (Table~\ref{table1}). In electron-doped layers, the slightly elongated \ce{Ru-O} bond indicates reduced covalency, while the filling of the minority-spin $a_{1g}$ band accounts for a moderate reduction in the spin moment. In contrast, in hole-doped layers, increased \ce{Ru-O} covalency, due to shorter bond distances, combined with depopulation of the majority-spin $e'_{g}$ band, results in a significant reduction of the \ce{Ru} spin moment.

Secondly, the SO coupling is influenced by charge doping, primarily through changes in the electronic structure. In electron-doped \ce{Ru2O6} sublayers, SO coupling is enhanced, as evidenced by an increase in the orbital moment from $m_{\rm o | Ru} \sim 0.04$ to $0.08 \mu_\text{B}$ (Table~\ref{table1}). Conversely, a reduction in the orbital moment is observed in hole-doped sublayers. Furthermore, the spin and orbital moments align parallel in electron-doped cases as the $t_{2g}$ band becomes more than half-filled.

\begin{figure}[!t]
\includegraphics[width=0.99\linewidth]{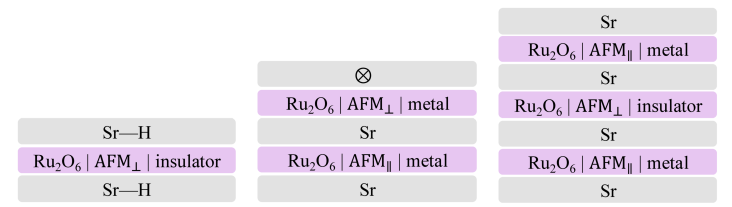}
\caption{Schematic illustrations of electronic and magnetic states in selected nanosheets reveal that the magnetic \ce{Ru2O6} sublayers consistently exhibit N\'eel AFM ordering, regardless of doping.  However, the magnetic anisotropy switches from an out-of-plane direction (AFM$_\perp$) to an in-plane direction (AFM$_\parallel$) in electron-doped sublayers. Additionally, charge carrier doping induces an insulator-to-metal transition within the \ce{Ru2O6} sublayers. These naturally grown nanosheets with intrinsic charge doping feature complex combinations of layered electronic and magnetic states that hold promising potential for spintronic applications.  
}
\label{fig:ComplexStack}
\end{figure}

Lastly, charge doping significantly alters the on-site magnetic anisotropy (Table~\ref{table1}). In neutral \ce{SrRu2O6} nanosheets, magnetism is characterized by perpendicular anisotropy (AFM$_\perp$, $A_z > 0$) relative to the 2D honeycomb \ce{Ru} lattice, as illustrated in Figure~\ref{fig:ComplexStack}. However, electron doping induces a transition to in-plane anisotropy (AFM$_\parallel$, $A_z < 0$) in the \ce{Ru2O6} sublayer. In contrast, hole-doped sublayers retain their easy-axis AFM$_\perp$ magnetism, preserving the perpendicular alignment (Figure~\ref{fig:ComplexStack}). The transition in the direction of magnetic anisotropy can be explained by treating SO coupling within the second-order perturbation theory.~\citep{PhysRevB.47.14932} First, the matrix element $\langle d_{xy}| L_z | d_{x^2-y^2}\rangle$, which promotes in-plane magnetic anisotropy, increases significantly in the electron-doped \ce{Ru2O6} sublayer. Second, the orbital-projected contribution from $\langle d_{z^2}| L_x | d_{yz}\rangle$, typically favoring out-of-plane anisotropy, changes sign in the electron-doped sublayer, as both orbitals predominantly host minority spins. These changes collectively drive the observed transition from perpendicular to in-plane magnetic anisotropy.

The bilayer and trilayer flakes exhibit intricate combinations of electronic and magnetic states across adjacent \ce{Ru2O6} sublayers, separated by atomically thin insulating \ce{Sr} layers (Figure~\ref{fig:ComplexStack}). For example, the [\ce{Sr}$|$$\cdots$$|$\ce{Sr}] trilayer demonstrates a [metal-AFM$_\parallel$ $|$ insulator-AFM$_\perp$ $|$ metal-AFM$_\parallel$] configuration. The extreme misalignment of magnetic anisotropy between adjacent layers can be exploited to harness field-induced phenomena with promising applications.~\citep{s41567-018-0050-y} In doped bilayer nanosheets (Figure~\ref{fig:ComplexStack}), the N\'eel vectors in the 2D metallic \ce{Ru2O6} layers are mutually orthogonal ($\varphi = \ang{90}$), separated by an insulating \ce{Sr} spacer layer. This configuration enables the bilayer nanosheets to act as antiferromagnetic tunnel junctions,~\citep{PhysRevB.73.214426} yielding a high-resistance state in the absence of an external magnetic field. Applying a magnetic field along the easy (hard) axis of the AFM$_\perp$ (AFM$_\parallel$) monolayer gradually reorients the N\'eel vector in one of the layers. At sufficiently high field strengths, the N\'eel vectors in both \ce{Ru2O6} layers align parallel ($\varphi = \ang{0}$), resulting in a low-resistance state. This tunability suggests that such exfoliated flakes could be integrated into magnetoresistive devices, similar to the behavior observed in orthogonally twisted ferromagnetic \ce{CrSBr} monolayers.~\citep{s41563-023-01735-6,s41586-024-07818-x,2410.19209} Additionally, the ability to manipulate the magnetic field strength and direction offers opportunities to engineer complex spin textures, providing further avenues for functional magnetic applications. Alternatively, the N\'eel vectors can be tuned electrostatically by gating, which drives the magnetic layers from the electron-doped regime to the neutral or hole-doped regime. 

\begin{figure}[!t]
\includegraphics[width=0.85\linewidth,trim={0 0 0 0},clip]{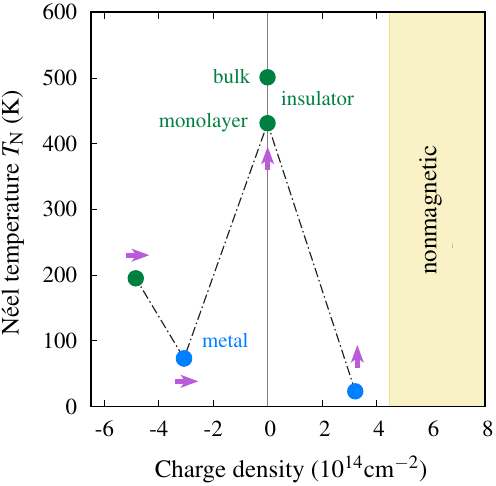}
\caption{The calculated N\'eel temperature exhibits nonmonotonic behaviour with charge carrier density as shown for the monolayer. The \ce{Ru} moment, exchange interactions, and magnetic anisotropy are altered in charge-doped nanosheets (Table~\ref{table1}). The calculated bulk $T_\text{N}$ aligns well with experimental results,~\citep{AngewChem53_4423_2014_Hiley,PhysRevB.92.104413_Hiley,PhysRevB.92.100404_Tian,acs.chemmater.0c02469} while the ordering temperature is renormalized in the monolayer. Charge doping is found to be detrimental to magnetic ordering, with a pronounced effect observed in hole-doped samples. Furthermore, the magnetic anisotropy, indicated by the arrow, transitions to the in-plane direction ($\rightarrow$), aligning parallel to the honeycomb \ce{Ru} lattice.
}
\label{fig:Fig.TN}
\end{figure}

To gain further microscopic insights into the magnetism of charge-doped flakes, we calculate the exchange interactions, which are significantly altered by the added electrons or holes to the \ce{Ru2O6} sublayers (Table~\ref{table1}). Electron doping leads to a monotonic reduction in $J_1$ with increasing carrier density, accompanied by drastic changes in the second-neighbor interaction $J_2$. In [\ce{H-Sr}$|$\ce{Ru2O6}$|$\ce{Sr}] flake, the AFM $J_2$ significantly increases, with $J_2/J_1 \sim 0.19$, compared to the negligible $J_2/J_1 \sim 0.01$ observed in the passivated monolayer. Upon further electron doping, $J_2$ transitions to a ferromagnetic interaction in [\ce{Sr}$|$\ce{Ru2O6}$|$\ce{Sr}] nanosheet. Furthermore, SO coupling is enhanced with increasing electron density as the $t_{2g}$ occupancy surpasses half-filling, leading to a pronounced increase in in-plane magnetic anisotropy. Using these exchange parameters, we compute the  N\'eel temperatures via Heisenberg Monte Carlo simulations. Notably, $T_\text{N}$ exhibits nonmonotonic behaviour (Figure~\ref{fig:Fig.TN}), driven primarily by the changes in the nature of $J_2$. Since the N\'eel AFM configuration on the honeycomb lattice is stabilized by FM $J_2$ and AFM $J_3$, variations in exchange interactions predominantly dictate the reduction in $T_\text{N}$. This effect is further compounded by a moderate quenching of the \ce{Ru} spin moment, which decreases to 1$\mu_\text{B}$. As expected, the exchange interactions in the electron-doped [\ce{Sr}$|$$\cdots$$|$\ce{Sr}] bilayer closely resemble those of the [\ce{H-Sr}$|$\ce{Ru2O6}$|$\ce{Sr}] monolayer, resulting in a comparable ordering temperature.

The microscopic behavior in hole-doped nanosheets differs significantly (Table~\ref{table1}). Analysis of the exchange interactions in both monolayer and bilayer nanosheets reveals key insights into their ordering temperature. A moderate increase in AFM $J_1$, combined with a substantial FM $J_2$ ($J_2/J_1 \sim -0.25$), should theoretically stabilize the N\'eel AFM configuration. However, a large FM $J_3$ ($J_3/J_1 \sim -0.45$) and a significant quenching of the \ce{Ru} spin moment to 0.3$\mu_\text{B}$ primarily drive the magnetic ordering to much lower temperatures. As a result, the hole-doped monolayer and bilayer exhibit magnetic ordering below 45 K (Figure~\ref{fig:Fig.TN}), which is substantially lower than their charge-neutral counterparts. With further increases in hole density, the \ce{Ru} spin moment is fully quenched, leading to a non-magnetic state.

\section{Summary}
We present a comprehensive investigation of two-dimensional magnetism in atomically thin, non-vdW oxide \ce{SrRu2O6} sheets using first-principles calculations combined with classical Heisenberg Monte Carlo simulations. This study is further inspired by the recent successful synthesis of ultrathin \ce{SrRu2O6} nanosheets, down to three layers, accompanied by detailed structural characterizations.~\citep{ACS_ApplNanoMat4_9_9313_2021_Suvidya_SrRu2O6} Within the adopted approach, the bulk properties are accurately reproduced, exhibiting G-type AFM ordering with a high N\'eel temperature. Charge neutral flakes exhibit magnetic ordering with $T_\text{N} > $ 430 K, positioning them as the highest known two-dimensional antiferromagnets. Although the ordering temperature in the monolayer limit is reduced due to enhanced thermal spin fluctuations, the renormalization is minimal, with $T^\text{1L}_\text{N} \sim 0.86T^\text{bulk}_\text{N}$, significantly outperforming vdW magnets in retaining robust magnetic ordering at reduced dimensions.

Charge carrier doping profoundly influences the properties of \ce{SrRu2O6} nanosheets by altering their electronic structure, spin and orbital moments, spin-orbit coupling, exchange interactions, and on-site anisotropy. A charge-induced insulator-to-metal transition is observed, while robust antiferromagnetic ordering persists in these ultrathin sheets. This behavior contrasts with \ce{Na2IrO3} nanosheets, where charge doping induces a magnetic phase transition.~\citep{Roy_PhysRevB.110.L020403} The N\'eel temperature decreases with charge doping, with a more pronounced reduction in hole-doped nanosheets. A detailed microscopic model supports these observations, linking the changes in exchange interactions and spin moments to the observed reduction in $T_\text{N}$. The remarkable electronic and magnetic properties, including above-room-temperature magnetic ordering in the as-grown nanosheets, highlight their potential for antiferromagnetic spintronic applications. These findings warrant further theoretical and experimental exploration to unlock their full potential.

\section{Acknowledgements}
We would like to express our gratitude to Ashna Bajpai for valuable discussions, which began with the exfoliation of \ce{SrRu2O6} flakes in her group and ultimately led to the initiation of this theoretical investigation. We sincerely acknowledge the support and resources provided by the PARAM Brahma Facility at the Indian Institute of Science Education and Research, Pune, under the National Supercomputing Mission of the Government of India. Additionally, we acknowledge the funding from the National Mission on Interdisciplinary Cyber-Physical Systems (NM-ICPS) of the Department of Science and Technology, Government of India, through the I-HUB Quantum Technology Foundation, Pune, India. D.K.R. acknowledges the Council of Scientific and Industrial Research (CSIR) India for support through a research fellowship.


%

\end{document}